\documentclass[sigplan,screen]{acmart}
\usepackage{listings}
\usepackage{xcolor}
\usepackage{algpseudocode}
\usepackage{algorithm}
\usepackage{xspace}

\newcommand{\sys}{XProv\xspace}

\AtBeginDocument{%
  \providecommand\BibTeX{{%
    \normalfont B\kern-0.5em{\scshape i\kern-0.25em b}\kern-0.8em\TeX}}}

\setcopyright{acmcopyright}
\copyrightyear{2018}
\acmYear{2018}
\acmDOI{XXXXXXX.XXXXXXX}

\acmConference[Conference acronym 'XX]{Make sure to enter the correct
  conference title from your rights confirmation emai}{June 03--05,
  2018}{Woodstock, NY}
\acmPrice{15.00}
\acmISBN{978-1-4503-XXXX-X/18/06}

\begin{document}

\title{Learning Lineage Constraints for Data Science Operations [Vision]}


\author{Jinjin Zhao}
\email{j2zhao@uchicago.edu}
\affiliation{
  \institution{University of Chicago}
  \city{Chicago}
  \state{IL}
  \country{USA}
}


\begin{abstract}
Data science workflows often integrate functionalities from a diverse set of libraries and frameworks. Tasks such as debugging require data lineage that crosses library boundaries.
The problem is that the way that ``lineage'' is represented is often intimately tied to particular data models and data manipulation paradigms.
Inspired by the use of intermediate representations (IRs) in cross-library performance optimizations, this vision paper proposes a similar architecture for lineage -- how do we specify logical lineage across libraries in a common parameterized way?
In practice, cross-library workflows will contain both known operations and unknown operations, so a key design of \sys to link both materialized lineage graphs of data transformations and the aforementioned abstracted logical patterns. We further discuss early ideas on how to infer logical patterns when only the materialized graphs are available. 
\end{abstract}

\ccsdesc[500]{Computer systems organization~Embedded systems}
\ccsdesc[300]{Computer systems organization~Redundancy}
\ccsdesc{Computer systems organization~Robotics}
\ccsdesc[100]{Networks~Network reliability}




\maketitle

\section{Introduction}
A single data science project can incorporate a variety of languages, libraries, and data container models~\cite{palkar2017weld}. 
There are often different requirements for code-iteration speed, data processing volume, and performance optimizations at each step of the process \cite{notebooks}. 
The Python data science stack is especially inter-operable, allowing users to quickly integrate the benefits of the appropriate tool for the task.
However, it is well-known that bugs arise at library boundaries~\cite{sculley2015hidden}.
For example, moving from a pandas \texttt{DataFrame} to a numpy \texttt{array} replaces column names with integer indexes, and mistakes can arise if a developer incorrectly maps indexes to column names.
This type of bug underscores the importance of ``cross-library'' data lineage, where data movement can be tracked and queried across the entire workflow.

Effective data lineage can enable applications in data privacy, data fairness, performance optimization, and code validation \cite{mlinspect, magic_push, modin, macke2020fine, psallidas2018smoke, glavic2021data, chapman2020capturing, phani2021lima, tang2019sac}. 
However, the representation of data lineage is tied to the underlying data model and data manipulation paradigm.
For example, in a relational workflow, storing predicate conditions is enough to represent a filtering operation since the data don't have a particular ordering~\cite{magic_push, modin, psallidas2018smoke}, but, in a DataFrame or Array workflow, ordering must be represented.

In such a cross-library environment, no good lineage options exist.
One extreme is to track lineage at a dataset level, i.e., which datasets are consumed by a workflow, where such issues of data representation are simply avoided~\cite{namaki2020vamsa}.
The other extreme is to track lineage at a Python variable level and avoid imputing semantics to those variables~\cite{macke2020fine}.
While such low-level tracking is the most general approach to lineage capture it comes with a steep overhead in both acquisition (instrumenting a library to track variables at a memory level) and storage (materializing a lineage graph of these variables).
It would be preferable to encapsulate common, known functions with a simple logical description also called logical lineage.
However, in a cross-library environment, there isn't an established language for these logical descriptions.

Inspired by the use of intermediate representations (IRs) cross-library performance optimizations~\cite{palkar2017weld}, this vision paper proposes a similar architecture for data lineage called \sys.
Namely, how do we specify logical lineage in a cross-library environment in a common parameterized way?
The IR that we propose is simple: (1) we settle on a unified data model of multi-dimensional arrays, and (2) library function calls are annotated with the types of lineage relationships they could possibly generate.
We call this IR a ``lineage-constraint tags'', as it constrains the type of input-output pairs that could be seen in a lineage graph generated by the operation.
As with any IR, there will be a combination of abstractable common user-defined patterns of lineage and black-box operations that are treated as materialized lineage graphs.
Even when incomplete, these additional semantics over the lineage graph introduce opportunities for storage and performance optimization of lineage. They further allow for much more effective queries across library boundaries. 
\emph{In short, \sys augments low-level physical lineage tracking with a higher-level intermediate provenance representation.}

The proposal is similar in spirit to verified lifting~\cite{cheung2013optimizing}, where operations in a low-level language are translated into a platform-agnostic higher-level representation.
Similarly, we take memory-level lineage captures from different libraries and impute higher-level operational semantics.
The interesting opportunity is that these operational semantics don't have to be perfect to be useful. 
This general approach also forms the basis for early ideas on learnt lineage capture. To support lineage over unknown operations, we propose that the patterns represented by lineage constraint tags can be learnt using recurring calls to the same operation, and that these patterns can be extrapolated to a new call on a new data container. In this manner, it is possible to capture the lineage of this operation, without exposure to any intermediate data states, and without the underlying modifying structure.


\subsection{Example Applications}
\begin{figure}[h]
\centering
\begin{tabular}{c}
\textbf{(A) Example Data Science Pipeline} \\ 
\includegraphics[width=.8 \linewidth]{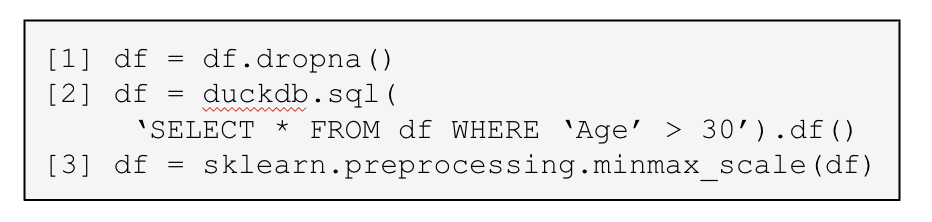}
\end{tabular}
\begin{tabular}{c}
\textbf{(B) Example Initial Dataframe} \\ 
\includegraphics[width=.4 \linewidth]{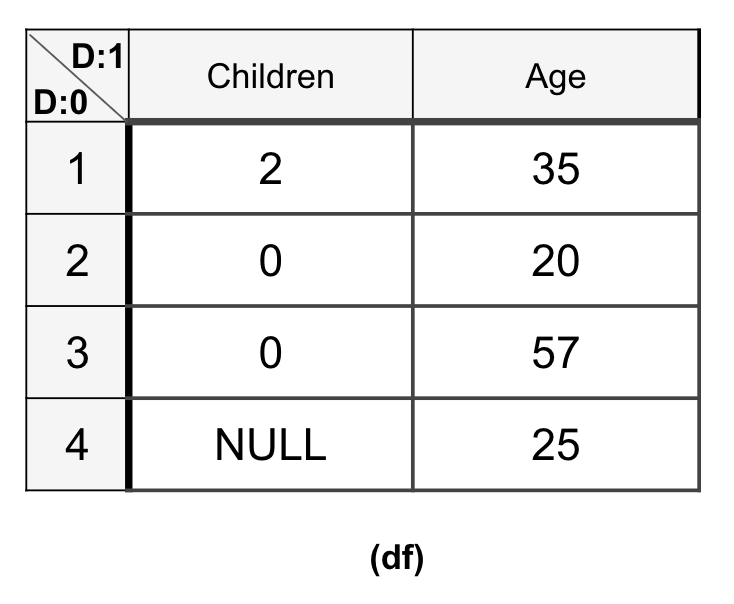} 
\end{tabular}
\vspace{-5pt}
\caption{We show (A) a simple example pipeline where a Pandas, DuckDB, and scikit-learn operations are applied to (B) a sample dataframe.}
\label{fig:example_code}
\end{figure}
Figure \ref{fig:example_code} outlines a simple data science pipeline. In this pipeline (A), the dataframe \texttt{df} (B) goes through a typical series of operations: (1) rows with \texttt{Null} values are removed using the Pandas library, (2) rows are selected where \texttt{`Age' > 30} with DuckDB, (3) columns are scaled between [0, 1] using \texttt{scikit-learn}. Note that, even though the pipeline is written in only Python, it interacts with three libraries that each have different data access patterns, and different scopes of operations. We present two applications over this pipline that we envision \sys to solve.

\begin{figure}[h]
\centering

\begin{tabular}{c}
\textbf{(A) Testing Dataset Leakage} \\ 
\includegraphics[width=.8 \linewidth]{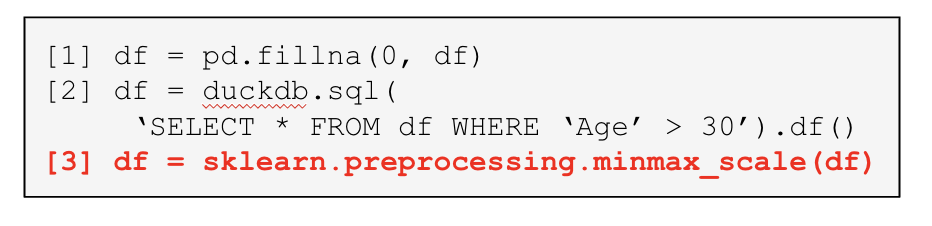}
\end{tabular}
\begin{tabular}{c}
\textbf{(B) Valid Execution Reorder} \\ 
 \includegraphics[width=.8 \linewidth]{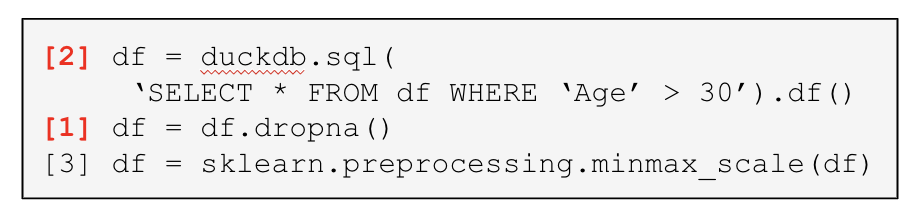} 
\end{tabular}
\caption{Lineage applications can apply information leakage checks (A) and operation reordering (B) to the example pipeline.}
\label{fig:example_applications}
\end{figure}

\textbf{Information Leakage.} Information leakage is a significant issue in machine learning pipelines. A typical scenario of leakage is when testing data is accidentally used in machine learning model training \cite{leakage_science}. We envision that this type of leakage can be addressed by using \sys to track data lineage in pipelines. In our example scenario, the dataframe has not been split into training and testing datasets. In this case, to prevent information leakage, we could enforce that all operations can only be ``row-wise''\cite{leakage_ml}. \sys can enable such enforcement, by directly answering queries about operation lineage. As seen in Figure \ref{fig:example_applications} (A), over the example code, \sys can notify a downstream application that the scaling operation is not ``row-wise'', and the downstream application can highlight the operation's code in red for users to modify.

\textbf{Performance Optimization.} Data operations are usually expensive at scale. One method of optimizing these operations is changing the order of execution, as seen in query schema rewrites \cite{magic_push}. As observed in Figure \ref{fig:example_applications} (B), in our example, one possible program rewrite is to switch the order of operations between the DuckDB operation, and the Pandas operation. This might improve performance significantly if the DuckDB operation is more selective. However, such rewrite is only feasible if we know that both operations are performing row-slicing on the dataframe, and switching their order does not change the output. Given that \sys could capture this information, we can enable downstream applications to apply ``rule-based'' re-writes across frameworks \cite{rewrite}.

\section{Proposed System Architecture}
We start by describing a system architecture for representing cross-library data lineage.

\subsection{Array Data Model}
\sys all significant data in a data science workflow as multidimensional arrays (called ``data containers'' in the paper to avoid confusion with original data type). The goal of this data model is to be deliberately broad to encapsulate a wide range of data such as dataframes, relational tables, spreadsheets, arrays, and tensor. This allows lineage to be tracked across different data framework within the life cycle of a data science project.

\sys reduces the native data type to arrays by determining the number of dimensions and a set of unique index values for each dimension needed to appropriately represent the structure. These properties have natural representation in our supported data types. For example, in relational tables, the primary key and column names directly corresponds to the index values. In dataframes, there is inherently support for row and column indices. Sub-keys can be used to resolve duplicates. 

Some data containers (i.e. dataframes, spreadsheets, arrays) also have natural ordering along dimensions, and we would also like to track lineage with respect to those orderings. This could be captured by storing indices in an ordered set when applicable, and reordering when the data container is reordered correspondingly.

\subsection{Producing a Data Dependency Graph}
In the first step of \sys, we plan to analyze the code to produce a data dependency graph (DAG). This graph shows all variable assignments and the variables and function calls needed to resolve those assignments.
For Python, it is relatively straight-forward to produce such a graph using internal static analysis modules \cite{mlinspect}.
Since Python is an interpreted, dynamically typed language, we resolve those variables to types only during execution.
Thus, each node of the DAG represents a Python subroutine in the user-written code and these can be resolved to particular library functions and data types after execution. 

\begin{figure}[t]
\centering
\includegraphics[width=.9 \linewidth]{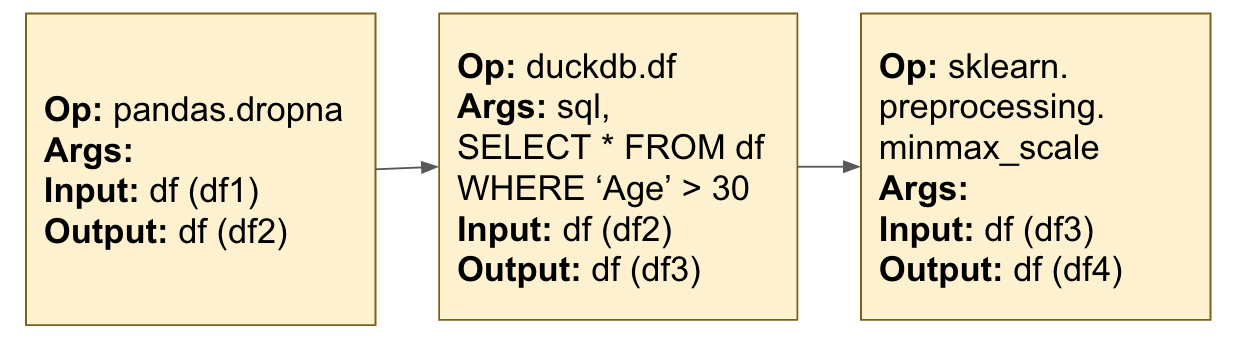} 

\caption{For the example code, we show the envisioned data lineage DAG stored in \sys.}
\label{fig:example_dag}
\end{figure}

Figure \ref{fig:example_dag} (A) shows an envisioned DAG with node signatures for our example pipeline. In this representation, we only focus on Python ingestion and the SQL query is treated as a parameter into the DuckDB operator. To the best of our knowledge, generating DAGs outside of Python (e.g. spreadsheets) is an open research problem.

\subsection{Lineage for Each DAG Node}

\begin{figure}[t]
\centering
\includegraphics[width=.7 \linewidth]{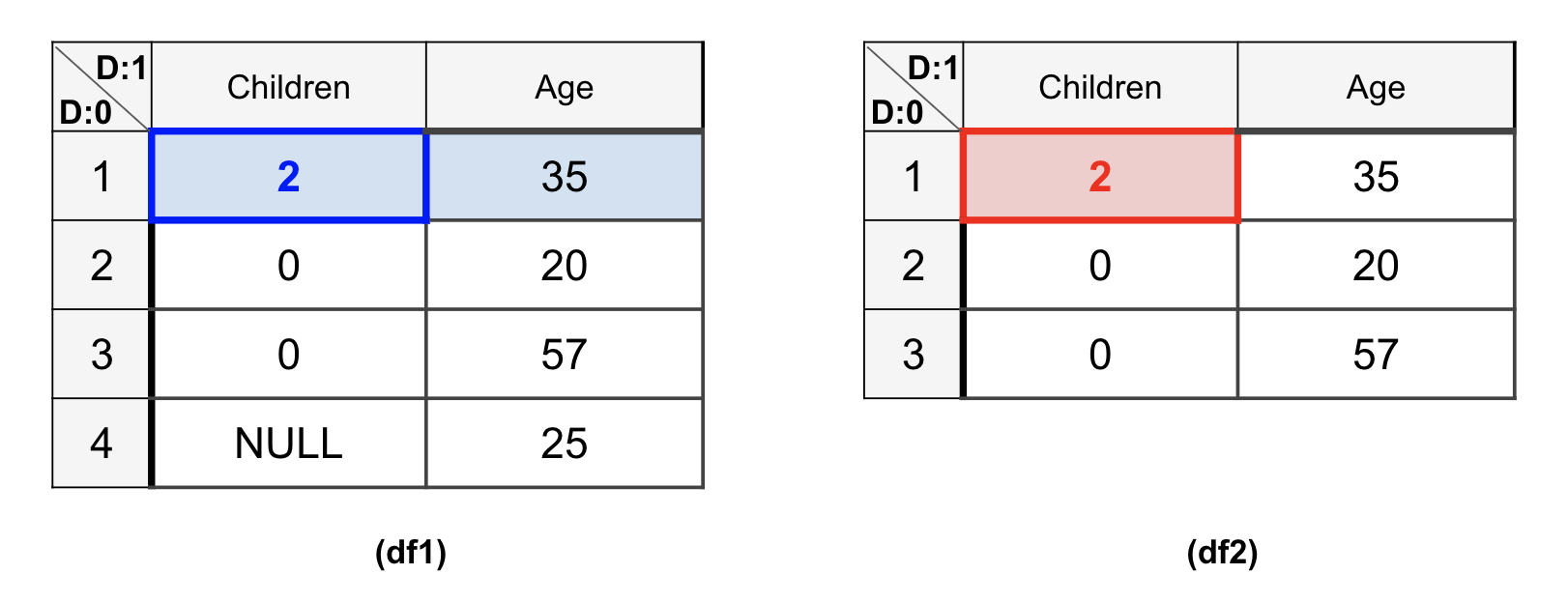} 

\caption{We show example input and output data containers for \protect{\texttt{\{OPERATION: pandas.dropna\}}}. In this example, the entity with blue text and the entities with light blue background  directly and indirectly influence the entity at [`Children', `1'], respectively.}
\label{fig:influence_examples}
\end{figure}

Previous work has catalogued different variations on the basic definition of why-provenance, as well as variations on how, for given data operations, to form a witness set of contributing input entities for an output entity \cite{provenance}. While we envision \sys would support multiple variations on why-provenance, in this paper, we focus on two particular forms that we call ``direct and indirect influence lineages''. These definitions of lineages not novel but correspond nicely with our early ideas on how to capture this lineage for operations where the underlying implementation and intermediate data states are unknown.

Suppose we have a operation signature, $\sigma$, that takes in an input data container, $A$, with a value at indices, $a$, and generates an output data container, $B$ with a value at indices, $b$. 

We defined $\{\mathbf{A}\}$ as the set of variations on the input data container, $A$, where $\forall A' \in \{\mathbf{A}\}, A'[a] = A[a]$ and $\sigma(A') = B$.  Now, let $\mathbb{A}$ be the domain of possible values for $A[a]$ given the context of the data container. If there exists $A' \in \{\mathbf{A}\}$, where for \textbf{any} value $x, x \neq A[a], x \in \mathbb{A}$ such that if we set $A'[a] = x$, $\sigma(A')[b] \neq B[b]$, then we state that the entity at index $a$ \textbf{directly influences} the entity at index $b$.  If there exists $A' \in \{\mathbf{A}\}$, where there exists \textbf{one} value $x, x \in \mathbb{A}$, such that when we set $A'[a] = x,$ $\sigma(A')[b] \neq B[b]$ or $b \notin \sigma(A')$, then we state that the entity at index $a$ \textbf{indirectly influences} the entity at index $b$. 

Let us see an example of these lineage concepts by considering the first operation signature in our example pipeline: \texttt{\{OPERATION: 'pandas.dropna'\}}. Figure \ref{fig:influence_examples} shows the input and output containers over this operation, as well as the lineage of the output entity at index [`Children', `1' ]. We can observe that this output is directly influenced by the input entity with the same indices, since its value is always the same. It is indirectly influenced by the input entities along the same row, since if any of those entities are `NULL', the entire row is removed from the output.

\sys will represent the influence lineages as a relational table using our prior work on lineage storage, DSLog \footnote{under submission.}. For each entity in the output container, we represent the relationship between its indices, and the indices of an influencing entity as a row in the table. This allows queries directly over the lineage relationships, as well as compression opportunities over the table itself. 

\subsection{Lineage-Constraint Tags}
\begin{table*}
\footnotesize
\centering
\begin{tabular}{|c|p{0.4 \linewidth}|p{0.2\linewidth}|c|}
\hline
\textbf{ Tag} & \textbf{Definition} & Relational Operation & Example Operation\\\hline
One-to-One & Each output entity is directly and indirectly influenced by exactly one input entity at the same indices. & PROJECTION & - \\\hline
Slice[DIM] & Each output entity is directly and indirectly influenced by only the input entities that have the same index at DIM.  The output indices for the input and output containers are exactly the same, except along DIM. & SELECTION, PROJECTION & (1), (2), (3) \\\hline
Identity & Each output entity is directly influenced by one input entity with the same value. & SORT, SELECTION, PROJECTION, DROP DUPLICATE
& (1), (2) \\\hline
Condition[DIM, INDEX] & All the output entities along the DIM dimension are indirectly influenced only by some (>1) input entities with index INDEX at DIM and input entities that have the same index at DIM. & GROUP-BY, SELECTION, JOIN, SORT & (2) \\\hline

\end{tabular}
\vspace{10pt}
\caption{This table shows example lineage-constraint tags that can be defined in \sys.}
\label{tab:semantic_tags}
\end{table*}

In this section, we introduce lineage-constraint tags. These tags are defined as names that correspond to an assertion over the linage information and input and output data containers of an operation. They inform the users of some constraining property on a node signature's lineage graph. Defining constraining properties rather than exact lineage patterns is a simple but crucial design decision that allows \sys to operate over unbounded operations.

Table \ref{tab:semantic_tags} shows some sample lineage-constraint tags - simple ``One-to-One'' and ``Condition'' tags, and relevant tags in \sys for our example applications. These tags are explicitly defined over the operation's linage and input and output values. They can be parameterized to match a specific property; for example, ``Slice'' intakes a parameter for the particular dimension that the ``slicing'' occurs over. The table also shows the common relational algebra operations and example operations that match each lineage constraint tag. 

\sys is designed to store these lineage-constraint tags, and map them to node signatures and operation signatures. A tag maps to an operation signature, if and only if every possible node signature that contains that operation signature satisfies the tag. These mappings allows users to query over particular constraints of a node's lineage, rather than directly over lineage tables, significantly increasing the usability of our system.

We envision that each lineage-constraint tag contains an assertion function that validates whether any lineage, input and output instance matches that constraint. Users can use those assertions to tag lineage properties over their DAG graph. In addition to the assertion function, the tag also may be linked to maximum constraint functions. These functions returns lineage tables that contain the maximum constrained lineage given that an input container satisfies the tag condition.



In the current design of our system, lineage-constraint tags and associated assertion functions and maximum lineage mappings are to be human-generated, but we would like to explore automatic generation in future work. The concept of lineage-constraint tags has similarities with previous ideas on how-provenance. How-provenance documents the process of how the input entities translate to output entities in an operation \cite{provenance}. Prior work has shown that algebraic structures (e.g. semi-rings), can represent and propagate relational provenance \cite{semiring}. Since \sys allows a more flexible set of lineage patterns, such algebraic structures is not defined; however, lineage-constraint tags capture similar concept, and describe how operations link input entities to output entities. Linking these tags over multiple operators is up to end application, and in some cases it may be easier working with the lineage graphs directly.


\subsection{User Workflow}
\begin{figure}[h]
\centering

\includegraphics[width=0.9\linewidth]{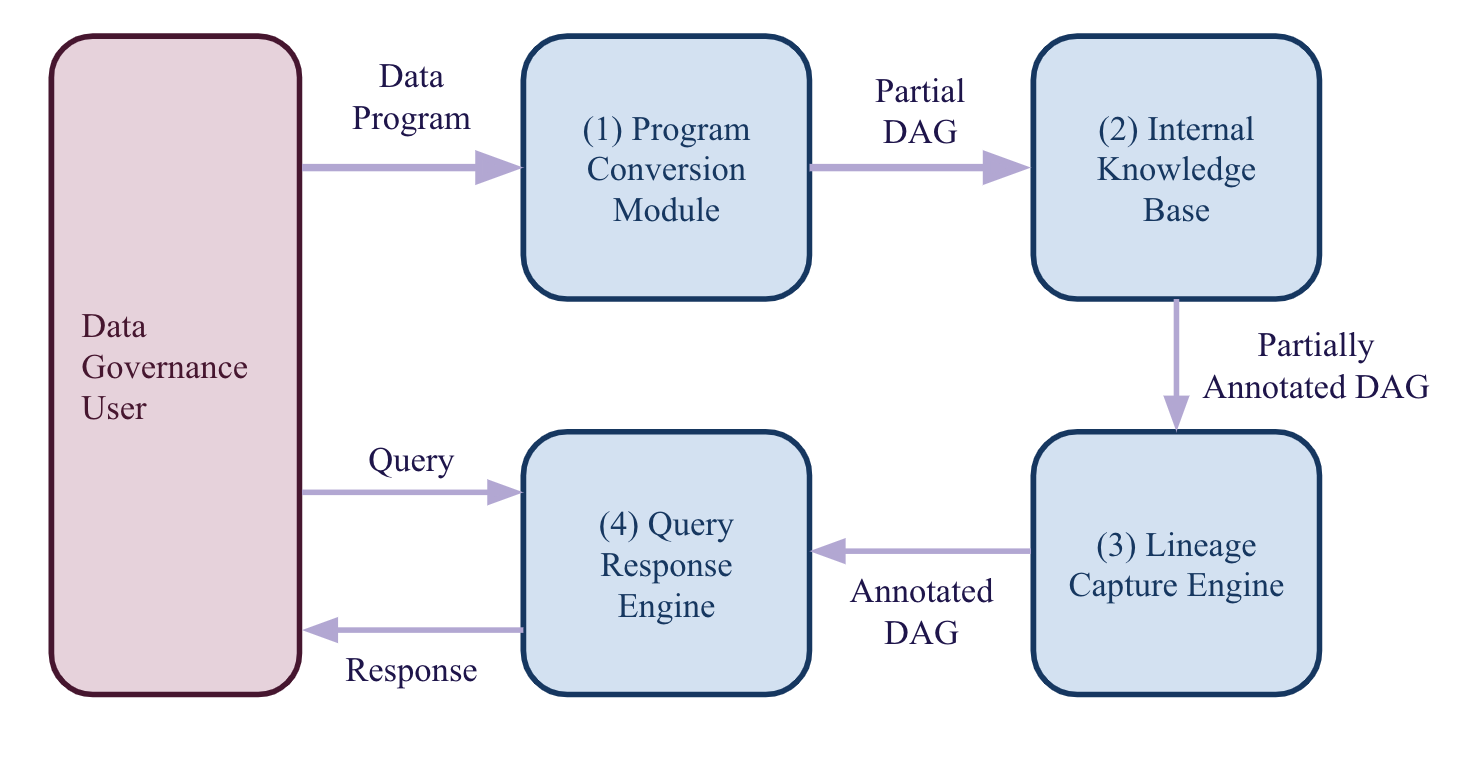}

\caption{We represent the architecture of \sys to enable (learnt) capture and query of lineage tables and semantic tags.}
\label{fig:architecture}

\end{figure}

Figure \ref{fig:architecture} summarizes the envisioned architecture of \sys. The system initially inputs a language-specific array-processing program. Within the system, the program is converted to an universal DAG dataflow by a language and framework specific module (1). Once the program is converted into the DAG, \sys queries its internal knowledge base for operation signature matches among its stored lineage tables and lineage-constraint tags (2). Since these properties may be learnt through various methods in the system, each of the matches contain origin logs on the action that generated the mapping between the signature and the lineage artifact. The users can consider this log along with their task-specific risk for lineage uncertainty. For example, we may accept learnt lineage tables for the information leakage application, but require exact lineage tables for changing the order of operation.

For operation signatures without matches, the system presents an option to learn lineage and lineage-constraint tags after execution of the graph, given that the operation signature itself is re-executable (3). For performance purposes, we aim to only execute the operation signature only once on the full data container.

Finally, the full lineage graph is available for querying for downstream tasks (4). \sys presents two APIs to enable queries over the lineage tables and semantic tags. Firstly, over any path along the initial DAG graph, \sys can query for lineage relationships between an input container and output container. The API for this query is presented below. This API returns \texttt{lin\_type} lineage between the entities indexed by \texttt{query\_indices} in the first container and relevant entities last container along the path given by \texttt{query\_path}.

\begin{lstlisting}[language=Python]
prov_query(query_path: Container_IDs [], query_indices: Indices, lin_type: String)
\end{lstlisting}
 
Secondly, for each node in the DAG graph, \sys can assert whether it contains particular semantic tags. The API call for assertions is shown below. This assertion can be internally executed by firstly searching over \sys's internal knowledge base and secondly applying the tag's assertion function over the operation signature's lineage table. If the optional \texttt{param} argument is filled, \sys can return whether a subset of parameters where the assertion is true.

\begin{lstlisting}[language=Python]
assert_tag(node: Node Signature or Operation Signature, tag: Semantic Tag, params: List:Optional)
\end{lstlisting}

\subsection{Example Applications}

Algorithm \ref{alg:applications} shows pseudocode representations of how \sys can be applied to our example applications. To evaluate potential information leakage, an application can apply the \texttt{row\_wise} function to all dataset operations before the training and testing split to ensure that no operation lineage connects across samples. \sys can also identify which operation performed the train-test split by tracking backwards lineage from the training and testing data container. For operations after the split, the application can use \sys to identify data leakages by evaluating whether there exists any lineage relations from the testing data container to the training container.

To discover potential candidates for execution re-ordering case, an application can apply the \texttt{double\_slice} function to parent-child node pairs, and assert whether both nodes slice the dataset along the same dimension. In this manner, additional rule-based execution optimizations can be evaluated by defining new semantic tags, and mapping operational signatures to those tags.

\begin{algorithm}
\caption{Using \sys's API for Lineage Applications}
\label{alg:applications}
\begin{algorithmic}
\Require $N \gets$ Node Signatures
\Procedure{row\_wise}{$N$}
\State $O \gets \{\}$
\ForAll{$n \in N$}
    \If{\texttt{!assert\_tag($n$, `Slice[0]')}} 
        \State $O.add(n)$
    \EndIf
\EndFor
\State \Return $O$
\EndProcedure
\Require $n \gets$ Parent Node Signature
\Require $m \gets$ Node Node Signature
\Procedure{double\_slice}{$n$, $m$}
\State $d \gets $\texttt{!assert\_tag($n_{op\_sig}$, `Slice', $n.DIM$)} 
\State $is\_id = $\texttt{!assert\_tag($n_{op\_sig}$, `Identity')} 
\If{$|d| \neq 0$ OR ! $is\_id$}
    \State \Return False
\EndIf
\State $d_2 \gets $\texttt{!assert\_tag($m_{op\_sig}$, `Slice', $d$)} 
\State $is\_id_2 \gets $\texttt{!assert\_tag($m_{op\_sig}$, `Identity')} 

\If{$|d_2 | = 0$ OR ! $is\_id_2$}
        \State \Return False
\EndIf

\State \Return True
\EndProcedure
\end{algorithmic}
\end{algorithm}

\section{Learnt Lineage Capture}
\begin{figure}[h]
\centering

\includegraphics[width=.6 \linewidth]{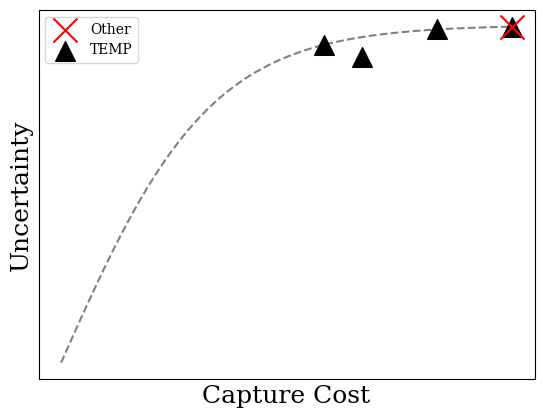}
\caption{\sys enables capturing lineage with uncertainty, which reduces the capture cost of lineage and increases the type of operation lineage that can be captured.} \label{fig:uncertainty}
\end{figure}

Suppose that there is an ``add one'' function that inputs one data container, and and adds one to all its entities. Theoretically, if this operation is a complete black box, it is impossible to determine the lineage for this operation, unless every possible input-output container pair is generated. This is easily shown with a proof-by-contradiction; for whatever input-output pair that is not generated, the operation can create a lineage that doesn't match the specification. However, after testing a varied examples set and observing that each output entity always is one more than the corresponding input entity, it would be a reasonable hypothesis to suppose that the function satisfies the "One-to-One" semantic tag. This example forms the basis of our idea that lineage can be learnt over black-box operations with reasonable uncertainty, and that this learnt lineage can useful for downstream tasks. To the best of our knowledge, previous lineage systems only supported high-cost lineage capture with certainty. \sys reduces the cost of capture when some uncertainty is allowed, and enables the capture of lineage for operations that were not feasible with complete certainty. 

\subsection{Lineage Generalization from Small Data Containers}
Lineage learning requires partial information into the black-box operation. There are many ways to gather this information. Some possible methods are probing a large language model to evaluate its knowledge of the operation, directly checking historical records of the input and outputs that \sys previously recorded, re-executing the full node signature with perturbations on the inputs, and testing small example inputs over the node's operation signature. 

In this paper, we focus on the last approach. This approach has particular properties that are attractive: (1) it doesn't require past records of the operation, and (2) it doesn't require multiple executions of the back-box node over large input containers. However, it does make the assumption that the operation signature is executable, and that lineage from small example inputs can generalize to larger inputs. In cases where the benefits are not necessary or the assumptions cannot hold, other methods could be explored. 

We propose early ideas on the algorithm for generating the example inputs to ingest into the operation signature. This algorithm is presented as a series of steps described below.  

\begin{enumerate}
    \item During the execution of the graph, the intermediate input container is recorded.
    \item A set of initial small containers is created by taking a subset of indices along each dimension of the container. If there are particular indices referenced in the operation signature, those indices are always preserved.
    \item For each container in the set, we generate a new set of containers by perturbing random individual elements of that container.
    \item The operation signature is applied to the generated small containers and the output containers are saved.
\end{enumerate}

Steps (1) and (2) are necessary so that the domain of the elements from the initial container is preserved. Step (3) draws directly from the definition of influence lineage. Under the assumption that the set $\{\mathbf{A}\}$ has a size of 1, any perturbations on an input element would affect all directly influenced output elements, and there would exist some perturbation that would affect all indirectly influenced output elements. Therefore, given enough perturbations, a downstream learning model would gain information about both the direct and indirect lineage of the given operation.

\subsection{Learning Semantic Tags and Lineage}
Now, let us briefly discuss how we can leverage machine learning to learn semantic tags and lineage from the given examples. Since semantic tags can be linked to operation signatures, they can sufficiently describe lineage constraints that persist over both the small data examples and the input data containers.

For each candidate tag, we plan to gather a positive and negative input-output pair examples. Based on these examples, and the set of small input-output pairs on the current operation signature, we envision that a machine learning model can return whether that operation satisfies each tag. Recent work on few-shot learning can be useful in this setting\cite{llm_learner} to minimize the number of previous example pairs stored.

Now, given a single tag, we can directly use its maximum constraint function to approximate the lineage of the black-box node. Given multiple tags, we know this learnt lineage can satisfy all constraints imposed by the tags. Therefore, the lineage approximation is just the intersection of lineage relationships outputted by all valid functions.
\section{General Discussion}
We highlight key observations about provenance in data science and how those lead to the design decisions made in \sys. 

\subsection{Data Manipulations Implementations are Heterogeneous.} 
In our past experience in provenance capture with DSLog, we found that low-level implementation of data science operations can be extremely heterogeneous. For example, most operations in the \texttt{numpy} library are based on low-level unary and binary operations in C. However, the linear algebra module uses BLAS and LAPACK packages that also directly modify the data in memory for better optimization. Another example is the Modin library \cite{modin}, which is based on the Pandas API, but changes the under-the-hood implementation to support parallel execution. 

Hence, it is extremely difficult to implement a provenance capture system based on the implementation of these operations. In this paper, we propose a vision for black-box provenance capture, where provenance is captured without knowing the underlying implementation of the operation, and without access to intermediate states of the data during operation execution. In the future, we would like to evaluate this capture method in terms of accuracy, coverage, and performance.

\subsection{Provenance Patterns are Consistent Across Data}
There are obvious similarities in provenance between large data structures and operations across frameworks. For example, practically all array-like data structures support a one-to-one map function, where a function is applied to each element in an array. Wu discuss similarities between the dataframe structure and the relational table \cite{Wu2019IsAD}, and Greene et. all. proposed semi-ring algebra to capture provenance patterns for database operations. \cite{semiring} By explicitly modeling both provenance and data structure types as separate objects in \sys, we can group the data structures by the provenance type and patterns that they support. In this paper, we introduce the concepts of provenance equivalence, and partial provenance matches. They allow the user to describe specific relationships between different types of provenance to reduce storage and compute costs, by using knowledge about one type of provenance to compute another type.

\subsection{Some Provenance Patterns are Easier to Capture than Others.}
To track the fine-grained provenance of the one-to-one map function, we only need knowledge about the shape of the data structure \cite{zhao2024compressioninsituqueryprocessing}. One the other hand, tracking fine-grained provenance of a filter function requires knowledge about the values of the input data structure \cite{wu2013subzero}. Additionally, the provenance pattern generated from the filter function is more complex, since there is greater data movement. There are implicit difficulty differences in tracking the provenance for different operations. Therefore, it may be feasibly to capture the fine-grained provenance of all the map functions in a workflow, but not the fine-grained provenance of the filter functions. To the best of our knowledge, \sys is the first system to explicitly define provenance graphs with partial information, and allow queries over those partial graphs. This is simply accomplished with the requirement of a null provenance value for each provenance type.

\subsection{Performance of Provenance Systems depend on End-Task.}
There are specific aspects of our system design made to optimize performance. By prioritizing storage of provenance information in the internal knowledge base, we reduce the cost of annotation and modification of the program itself when possible. 

Measuring the costs of \sys over different types of provenance would be a key evaluation. However, it is not quite clear-cut the targets that such costs need would need to match in order to be useful for such end tasks. For example, in the first case study, we might need to evaluate for information leakage at every modification of the machine learning pipeline. Therefore, we would need to ensure that the latency induced by the provenance tracking system doesn't break the flow in a user's thought process[?]. However, in the second case study, we might only need a schema rewrite optimization upon deployment of that program into production. In that case, the latency costs can be significantly higher.

\subsection{Integration of Large Language Models}
One paradigm shift in how data science workflows are developed is the recent rise in large language models (LLMs)\cite{attention}. These models have been proven to produce executable structured code from human-language descriptions for previously seen tasks, and reason about different data entities \cite{narayan2022foundationmodelswrangledata}. Given these facts, it is reasonable to propose that the LLMs have some capacity to reason about provenance. In this section, we discuss the impact of these models on \sys. We focus on the capacities of current models, noting that this field is nascent and fast-paced, and our discussion might not capture the state of such models in the future.

Our early experiments show that LLMs are good at provenance. We evaluated GPT-3.5 on example code of our two case studies and the example case study on fairness, as presented by Grafberger et al \cite{mlinspect}. For the task of capturing fine-grained provenance, GPT-3.5 was able to correctly reason about the provenance for singular functions, but was not about to track provenance across functions. For the task of schema rewrites, GPT-3.5 successfully re-ordered the two operations so that the projection occurred before the missing value filter when asked for performance optimizations. Finally, for the task of fairness, when presented with the healthcare example code, the model correctly identified that the selection operator might introduce unfairness. It also made correct suggestions that imputation, and inherently using certain features (such as ``income'') might lead to biases, which their tool did not suggest.

However, despite this positive result, there are known limitations of LLMs that suggest they cannot completely replace a transparent provenance system, such as \sys, in their current state. The most known negative quality of LLMs is that they ``hallucinate'' \cite{Huang_2025}. In other words, they sometimes present factually incorrect results in the exact manner and confidence of factually correct results. In sensitive tasks, such as creating data science pipelines for production, such errors might be expensive, and \sys presents a method of verification by transparently creating accurate provenance graphs.

Current LLMs are also presented as centralized services due to the resources required to operate such models. Even in the cases where the query to a LLM is usable, querying a local knowledge base for structured tasks, such as provenance generation, may have significant benefits in monetary costs, energy costs, and privacy protection. 

Finally, these models are presently limited to internal training data, and do not incrementally update their weights based on queries. Therefore, they may lack knowledge about niche operation signatures and data structures not readily present. For example, when asked about the Umbra database's internal query representation \cite{Neumann2020UmbraAD}, GPT-3.5 stated that it had no knowledge of Umbra, and Google Bard hallucinated an incorrect response.

We believe that LLMs can be integrated into \sys to provide better services on end tasks. Firstly, LLMs can be used as a capture method for the internal knowledge base. One drawback of Python in terms of provenance is that it does not require data types. Therefore, the data structures that are used as inputs into operations may not be explicitly defined. LLMs can make reasonable deductions on the data structures of well-known Python operations. Secondly, as shown above, LLMs seem to have reasonable accuracy in evaluating provenance for operations that were in its training dataset. For these operations, we can directly query for provenance from a LLM, and validate those queries with additional capture methods.

LLMs can also be users of provenance and systems like \sys. As mentioned above, \sys can provide verification on suggestions made by LLMs, and ensure that such suggestions are consistent with the provenance graphs generated. Additionally, these provenance graphs themselves may be used as inputs into LLMs to generate better suggestions on end tasks.

\section{Related Works}

\textbf{Data Quality Tools.} In recent years, there has been an acknowledgement that evaluating data quality is an important aspect of data science. A direct analogy can be made between data contracts and lineage constraint tags; data contracts check and enforce against some property on snapshots of individual datasets \cite{gable2023seed, soda2023contracts, montecarlo2023qa, datafold2022pr}. In a similar vein to learnt constraint tags, automatic generation of data contracts has been recently explored by the industry, where data contracts have been inferred by the underlying data structure.

\textbf{Lineage Systems.} Numerous existing end-to-end lineage systems ~\cite{xin2018helix, shang2109alpine, hellerstein2017ground, zaharia2018accelerating, derakhshan2020optimizing} focus on tracking coarse-grained lineage at the level of whole machine learning programs. While these systems are valuable for versioning and dataset discovery, they fall short in capturing fine-grained lineage and lack details on how individual array elements influence the final outcome. Systems that use lineage at a finer-granularity \cite{namaki2020vamsa, mlinspect, psallidas2018smoke, interlandi2015titian, modin, wu2013subzero} focus on specific languages and packages used in data science, rather than boundaries between those tools, and explicitly follow a particular lineage definition, without allowing uncertainty about lineage. 

\section{Author's Note}

The work presented here was a draft written in 2023 that extends ideas on provenance from the DSLog paper\cite{zhao2024compressioninsituqueryprocessing}. At the time, it was not submitted for publication, but after revisiting the paper, I found the ideas here interesting and relevant to current discussions on provenance with uncertainty.


\bibliographystyle{abbrv}
\bibliography{references}

\end{document}